\documentclass[12pt]{article}
\usepackage{PRIMEarxiv}
\usepackage{amsmath, amssymb}
\usepackage{natbib}
\usepackage{graphicx}
\usepackage{authblk}
\usepackage{hyperref}
\usepackage{array}

\newcommand{\RR}{\mathbb{R}}
\renewcommand{\vec}[1]{{\boldsymbol{\mathbf{#1}}}}
\begin{document}

\title{Efficient and robust control with spikes that constrain free energy}

\author[1]{André Urbano}
\author[1,2,*]{Pablo Lanillos}
\author[2,*]{Sander Keemink}
\affil[1]{Cajal Neuroscience Center, Spanish National Research Council, Madrid, Spain}
\affil[2]{Department of Machine Learning and Neural Computing, Donders Institute for Brain Cognition and Behaviour, Radboud University, Nijmegen, The Netherlands}
\affil[*]{These authors contributed equally.}


\maketitle

\begin{abstract}
Animal brains exhibit remarkable efficiency in perception and action, while being robust to both external and internal perturbations. 
The means by which brains accomplish this remains, for now, poorly understood, hindering our understanding of animal and human cognition, as well as our own implementation of efficient algorithms for control of dynamical systems.
A potential candidate for a robust mechanism of state estimation and action computation is the free energy principle, but existing implementations of this principle have largely relied on conventional, biologically implausible approaches without spikes. We propose a novel, efficient, and robust spiking control framework with realistic biological characteristics. The resulting networks function as free energy constrainers, in which neurons only fire if they reduce the free energy of their internal representation. 
The networks offer efficient operation through highly sparse activity while matching performance with other similar spiking frameworks, and have high resilience against both external (e.g. sensory noise or collisions) and internal perturbations  (e.g. synaptic noise and delays or neuron silencing) that such a network would be faced with when deployed by either an organism or an engineer.
Overall, our work provides a novel mathematical account for spiking control through constraining free energy, providing both better insight into how brain networks might leverage their spiking substrate and a new route for implementing efficient control algorithms in neuromorphic hardware. 
\end{abstract}


\noindent\textit{Correspondence:} andre.urbano@csic.es

\section{Introduction}
The human brain is capable of controlling more than 600 muscles to perform complex tasks requiring fine motor control, adaptation and higher-level cognition while expending only a few watts of energy through its spiking operations \cite{purves_neuroscience_2004}. 
Its internal components are biological, noisy and stochastic \cite{parnas_noise_1996, jacobson_subthreshold_2005}, yet it maintains remarkably robust control in the uncertain conditions of the natural world. 
This naturally leads us to a fundamental question: how to build control systems that exhibit similar robustness to internal perturbations (e.g., neuronal noise and component failures) and external perturbations (e.g., environmental uncertainty and sensory noise)?

At the computational level, one promising answer to adaptive behavior comes from the Free Energy Principle (FEP), which posits that organisms minimize a statistical quantity that bounds sensory surprisal~ \cite{friston_free_2006}. 
The mathematical framework behind this principle, Active Inference~\cite{parr_active_2022}, has been successful in explaining diverse phenomena, including perception, motor control, learning, and decision-making in both natural and artificial agents~\cite{lanillos_active_2021}. 
By treating both perception and action as Bayesian inference processes that minimize prediction errors about the world, Active Inference provides a unified computational account of adaptive behavior. 
However, despite its explanatory power at the computational level, there remains a critical gap: we lack bio-physically plausible mechanistic models that could actually implement Active Inference using the spiking dynamics observed in biological neural circuits.

On the other hand, Spiking Neural Networks (SNNs) have demonstrated considerable promise for implementing control systems with energy efficiency comparable to biological brains \cite{bing_survey_2018}. 
General-purpose frameworks such as Spike Coding Networks (SCNs) \cite{boerlin_predictive_2013, nardin_nonlinear_2021} and the Neural Engineering Framework (NEF) \cite{eliasmith_neural_2004} provide principled approaches to building spiking networks that can represent and transform information. 
While these frameworks have been successfully applied to control tasks \cite{slijkhuis2023closed, fuqiang_huang_optimizing_2017, huang_dynamical_2018} and exhibit robustness properties \cite{calaim_geometry_2022, ndri_predictive_2024}, they were designed as general computation substrates rather than being specifically optimized for control. 
This generality leads to computational inefficiencies with respect to their biological counterpart when imposing objective functions for adaptive control, such as the free energy. 
Instead of adapting these general frameworks, one cane derive a spiking controller from first principles by directly constraining free energy, inspired by the error-constraining mechanisms of SCNs but tailored specifically for Active Inference.

\begin{figure*}[hptb]
\centering
\includegraphics[width=0.99\linewidth]{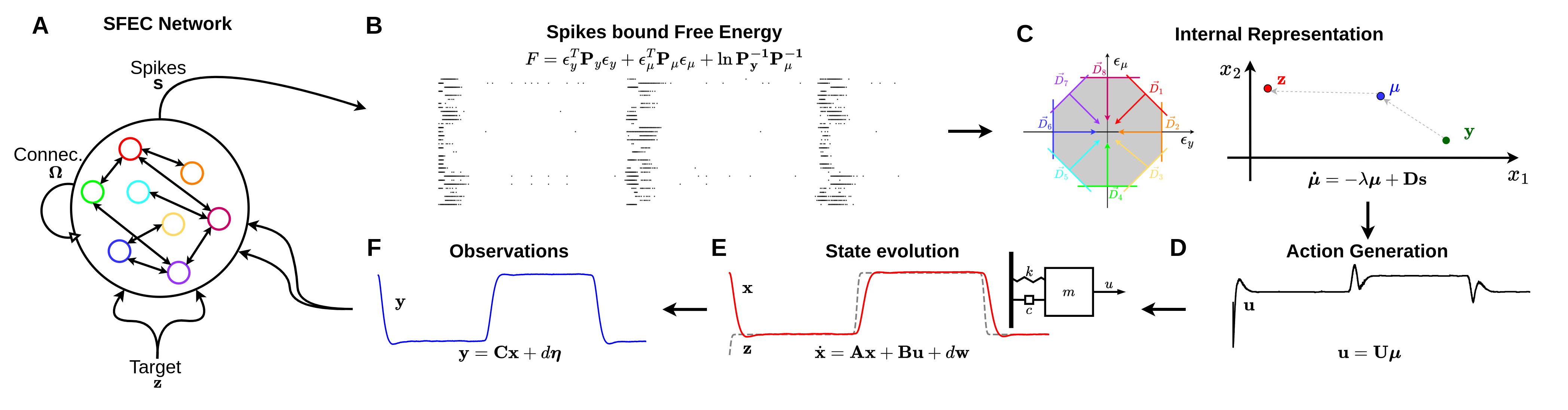}
\caption{Control Loop of Spiking Free Energy Constrainer (SFEC) networks, displayed schematically, for controlling a single Spring-Mass-Damper linear dynamical system.
A) The network receives the system observations ($\mathbf y$) and a target state ($\mathbf z$) as input, as well as recurrent connections of two different timescales ($\mathbf \Omega$), and outputs sequences of spikes that drive the network's internal representation.
B) The network's objective is to maintain a bounded free energy, which is computed from the precision-weighed prediction errors ($F$, top). This corresponds to a quadratic optimization process that maintains low prediction errors ~\citep{geirhos_shortcut_2020, mancoo_understanding_2020}.
Spikes are generated from this process during states with high free energy, as we can see by the fact that activity in the spike raster (bottom) is concentrated around the moments when the target changes instantaneously.
C) SFEC networks maintain an internal estimate ($\boldsymbol \mu$), which is driven by the spikes. The weights of the network are arranged such that the neurons' thresholds form a bounding box~\citep{calaim_geometry_2022} around the point of minimal free energy and each neuron's firing will return the signal closer to this point. The gray polygon within the box indicates the limits of where the internal errors are constrained to be ($\boldsymbol \epsilon_{\boldsymbol{\mu}}$ and $\boldsymbol \epsilon_{\boldsymbol{y}}$), thus constraining the free energy. When either error becomes to big, thus "hitting" one of the colored edges, the corresponding neuron fires, moving the errors along the vector $\boldsymbol{D_i}$ towards the center of the bounding box. The internal estimate is updated such that the free energy ($F$) remains bounded through a bounding box mechanism~\citep{calaim_geometry_2022}. Because the free energy combines measures of the error with the measured state and the target state, the internal estimate remains between the two.
D) Based on the internal estimate (which is biased towards the target state) a control action (force) is generated and applied to the system driving it from its actual state towards the internal state, ensuring it follows the target dynamics.
E) The state variables evolve according to the Plant dynamics, the control input and a process noise term. The system used in this example is the linear SMD system.
F) Observations are generated from the current state of the system and an observation noise term, and fed back into the network.}
\label{fig:Main}
\end{figure*}

Here, we propose and derive the Spiking Free Energy Constrainer (SFEC - Fig. \ref{fig:Main}A) by following the balanced network dynamics outlined in the SCN framework, but applying its error-constraining process directly to the free energy expression itself (Fig. \ref{fig:Main}B). 
Each spike updates our network's internal representation (Fig. \ref{fig:Main}C) and its action (Fig. \ref{fig:Main}D) to reduce free energy, naturally implementing Active Inference for control of linear dynamical systems while maintaining desirable robustness properties. 
Through systematic evaluation, we show that SFEC achieves comparable performance to similarly-sized networks of other frameworks and is robust to external kicks as well as process and observation noise. Importantly, the resulting networks expend orders of magnitude fewer spikes, exhibit graceful degradation under neuron silencing (maintaining good performance with 25\% neuron loss), and remain robust to voltage noise, spike transmission delays, and synaptic perturbations. These properties suggest SFEC as a promising neural model for adaptive robotic systems and energy-constrained neuromorphic platforms, while also providing a mechanistic bridge between the computational principles of Active Inference and the biophysical reality of spiking neural computation.

\section*{Results}
To build efficient spiking control systems that exhibit robustness to both internal and external perturbations,  we derive the Spiking Free Energy Constrainer (SFEC), a novel SNN that implements Active Inference by directly constraining the variational free energy through sparse spike emissions~\ref{fig:Main}. 
The network receives noisy observations of the system state (Fig.~\ref{fig:Main}F), fires spikes to adjust its internal estimate and constrain the free energy (Fig.~\ref{fig:Main}A-C), and produces the required control signal through a simple linear readout to move the real system state towards the internal state, which in turn moves towards the target state (Fig.~\ref{fig:Main}D-E). 
In Fig.~\ref{fig:Main} SFEC is applied to a simple Spring-Mass-Damper system. The real system state (red) is successfully moved towards a target (dashed line), using sparse spiking activity (raster plot in panel B). Note how spikes are mainly fired when the real state and internal state differ (resulting in high free energy), and are not required during state maintenance (since the free energy is low and few spikes are needed). This creates a natural correspondence between energy efficiency and control accuracy.

We will show that SFEC achieves three key properties: efficient control of diverse linear dynamical systems with minimal spike usage, flexible target dynamics that enable coordinated multi-agent behaviors, and robust performance under both external disturbances and internal network failures.
We first present the conceptual derivation of SFEC, then systematically evaluate its capabilities on various linear dynamical systems, from single-mass systems to coupled oscillators and coordinated drone swarms, and finally we quantitatively compare SFEC's efficiency and robustness against alternative spiking implementations.

\subsection*{Deriving the Spiking Free Energy Constrainer}
\label{Methods:SFEC}
Biological and artificial bodies can be described as dynamical systems. 
Consider a linear dynamical system with state $\mathbf{x}$ and dynamics:
$$\mathbf{\dot{x}} = \mathbf{A}\mathbf{x} + \mathbf{B}\mathbf{u} + d\mathbf{w}$$
$$\mathbf{y} = \mathbf{C}\mathbf{x} + d\boldsymbol{\eta}$$
where $\mathbf{A}$, $\mathbf{B}$ and $\mathbf{C}$ are the dynamics, input and observation matrices, and $d\mathbf{w}$ and $d\boldsymbol{\eta}$ are the corresponding process and observation noise terms.
Our objective is to compute an appropriate control signal $\mathbf{u}$ that drives the system toward a target state $\mathbf{z}$, while only having access to noisy observations $\mathbf{y}$.

Assuming that brain networks function as predictive systems~\cite{pezzulo_secret_2021} that minimize prediction errors~\cite{rao_active_2023}, we can frame this control problem through the lens of Active Inference. 
The basic idea is to maintain an internal state estimate of the system, incorporating the desired goal (perception), and to simultaneously generate the required control based on this internal state (action)~\cite{pezzato_novel_2020, lanillos_active_2021} (Fig.~\ref{fig:Main} C). 
Under the FEP~\cite{friston_free_2006}, this is implemented by minimizing the variational free energy, a function that balances two prediction errors: the difference between internal and desired states, and the difference between predicted and observed sensory inputs. 
Considering that states and observations are random variables that encode uncertainty, this prediction error minimization implements approximate Bayesian inference, making the internal belief distribution approximate the world state given observations. 
The variational free energy function $F$ in continuous form~\cite{lanillos_active_2021} can be expressed as
\begin{equation}
    F = \boldsymbol{\epsilon_y}^T \mathbf{P_y}\boldsymbol{\epsilon_y} + \boldsymbol{\epsilon_{\mu}}^T \mathbf{P_{\boldsymbol \mu}}\boldsymbol{\epsilon_{\boldsymbol \mu}} + \ln{\mathbf{P}^{-1}_{\mathbf{y}}} + \ln{\mathbf{P}^{-1}_{\boldsymbol{\mu}}}
    \label{eq:F}
\end{equation}
where $\boldsymbol \epsilon_\mathbf{y}$ and $\boldsymbol \epsilon_{\boldsymbol \mu}$ are the sensory and dynamics prediction errors respectively, and $P_y$ and $P_{\boldsymbol \mu}$ are their corresponding precisions or inverse variances ($\mathbf{P}^{-1}= \mathbf{\Sigma}$). 
The sensory prediction errors encode the distance between the observations predicted by our model and the real observations, whilst the dynamics prediction errors encode the distance between the position we believe we are in and the position we would like to be in, our ``target" position.
See Methods 
for a complete derivation. 
Minimization of the sensory prediction error ensures the internal estimate tracks the real state, while minimizing the dynamics prediction error ensures the internal state is biased towards the desired goal or target dynamics.

\paragraph{From gradient descent to spike-based free energy minimization.} 
The classic Active Inference approach in continuous time is to follow the negative gradient of the free energy function to find its minimum~\cite{pio-lopez_active_2016,sancaktar2020end, lanillos2020robot,oliver2021empirical}. 
While this can in principle be implemented in spiking networks through existing frameworks~\cite{eliasmith_neural_2004, boerlin_predictive_2013, isomura_canonical_2022}, such implementations are not optimized for Active Inference and do not necessarily leverage the sparse and efficient nature of spiking computation. 
For instance, SCNs which implement dynamical systems generate spikes to track an internal state, and can then evolve this internal state according to some target dynamics. One can then implement state dynamics that mimic the gradient descent from Active Inference, but the spiking itself is then not explicitly happening to reduce the Free Energy --- they are purely focused on representation.

Instead, here we derived a direct spiking solution by requiring that spikes directly consider the free energy, resulting in the Spiking Free Energy Constrainer (SFEC). Drawing inspiration from SCN theory~\cite{barrett_optimal_2016, deneve_brain_2017}, which is also strongly influenced by predictive coding, we derive network configurations and weights such that neurons only emit spikes when this would reduce the variational free energy: $F(\text{no spikes}) > F(\text{neuron spikes})$. 
This is equivalent to constraining the error terms underlying the free energy to be within some bound \citep{calaim_geometry_2022} — the neurons effectively draw a "bounding box" around the point of minimal error and spikes are fired when the internal errors grow too large (Fig. ~\ref{fig:Main}C). 
In effect, high network activity represents high free energy states that the controller seeks to avoid, whereas low activity implies the signal is close to its target position and only sparse maintenance spikes need to be emitted (Fig. \ref{fig:Main}B).

\paragraph{Network architecture and dynamics.}
From the above constraining principle, we derived a closed-form network of integrate-and-fire (IF) neurons for controlling dynamical systems [Ref Methods]~\ref{Methods:SFEC}. 
Assuming a linear readout of the network as a decoder of spike rates $\boldsymbol{\mu}=\mathbf{D}\boldsymbol{r}$, the internal voltage dynamics of neurons under the proposed framework have three types of input: input connections $\mathbf{W}_y$ that encode information about the real state of the system, fast timescale recurrent connections $\mathbf{\Omega}_{fast}$ that ensures the instantaneous internal network state ($\boldsymbol \mu$) is maintaned, and slow timescale recurrent connections $\mathbf{\Omega}_{slow}$ that encode the system's beliefs regarding its dynamics. 
The resulting voltage dynamics is:
\begin{equation}
    \dot{\vec{v}} = \mathbf{W}_y \dot{\vec{y}}^{+} + \mathbf{\Omega}_{slow}\vec{r} + \mathbf{\Omega}_{fast}\vec{s} 
\end{equation}
where $\vec{y}^{+}$ is a vector containing all inputs (target and observations), $\vec{r}$ are the neuron firing rates, and $\vec{s}$ are the spikes in binary format (spike=1, no spike=0).

\subsection*{SFEC controls single and coupled oscillatory systems}
\begin{figure}[htbp]
\centering
\includegraphics[width=0.8\linewidth]{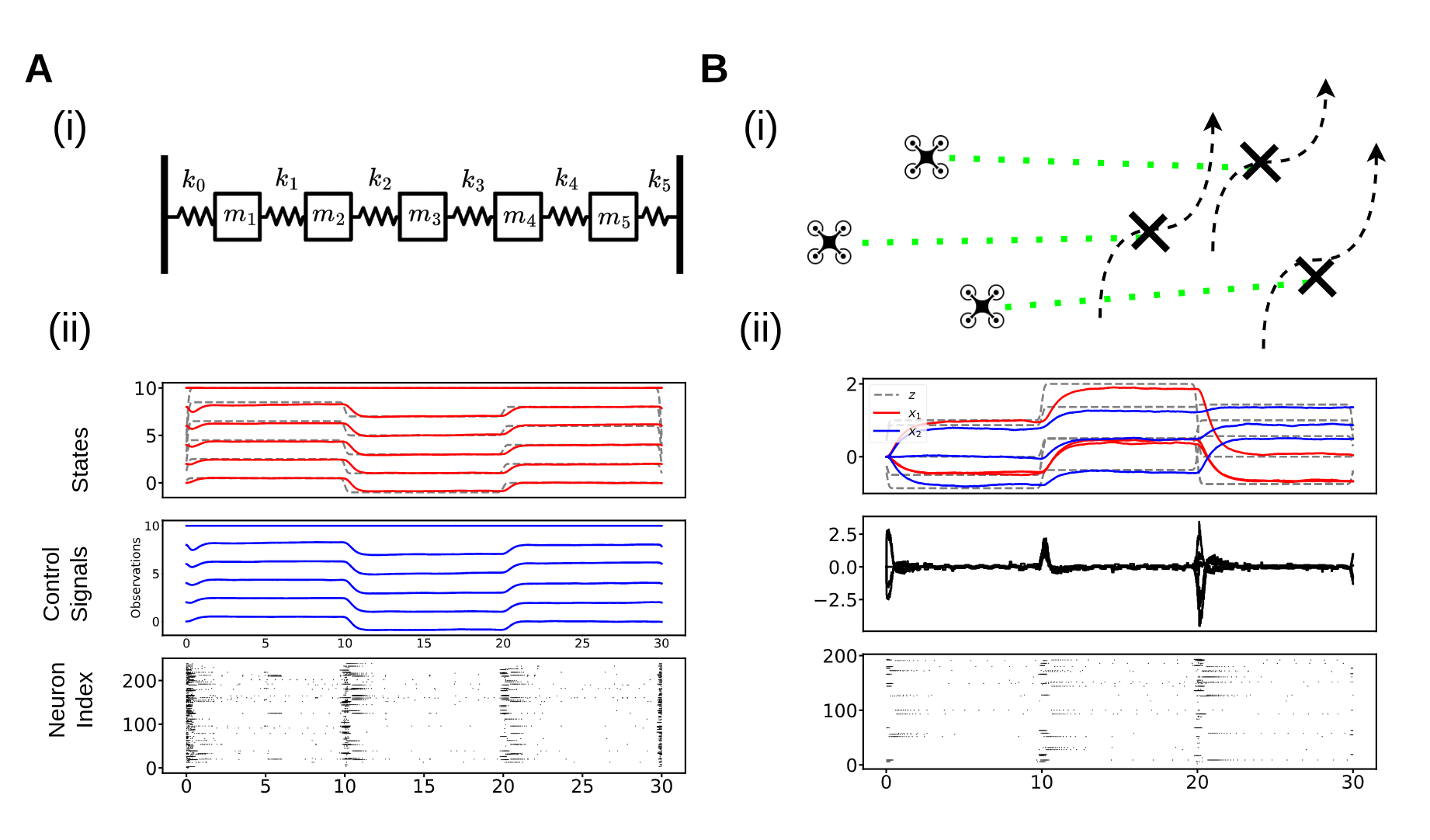}
\caption{SFEC can successfully control coupled dynamical systems.
A) (i) Schematic of a Coupled Oscillator system with 5 masses and 6 springs. The masses oscillate between two walls with fixed positions at $x=0$ and $x=L$. (ii) Control of a coupled Oscillator system showing the oscillator states converging to the target states (top), the control signals generated by the control algorithm (middle), and the spike raster showcasing network activity (bottom). B) (i) Schematic of a system of drones (modelled as 2D free masses). The green lines represent the attractive forces the controller exerts towards the target positions, the big X's are those target positions and the dotted arrows are the trajectory the target signal is following (ii) Same as panel A (ii) but for control of a drone swarm system.
}
\label{fig:CoupledOscillatorAndDrones}
\end{figure}

\paragraph{Problem:} Can SFEC effectively control physical systems with realistic dynamics? 
We begin with two fundamental test cases: a single spring-mass-damper (SMD) system and a coupled oscillator system, where multiple masses are connected by springs (Fig. \ref{fig:Main}E, Fig. \ref{fig:CoupledOscillatorAndDrones}Ai). 
These systems represent common control challenges — stabilizing oscillatory dynamics and managing coupled interactions — while allowing us to verify SFEC's basic functionality under noise.
\paragraph{Approach:} 
The Spring-Mass-Damper system consists of a single mass m attached to a fixed point by a spring of spring constant k and experiencing damping force with coefficient c.
The Coupled Oscillator system extends this to n+1 masses in series, each with mass m, connected by springs of different spring constants $\vec{k}$ and experiencing damping force c (Fig.~\ref{fig:CoupledOscillatorAndDrones}A i). 
The final mass is fixed in position. 
In both systems, the controller actuates along velocity dimensions by outputting forces. 
Controller parameters were manually tuned to suppress ringing oscillations. We provide target states that generally remain stable, but occasionally change suddenly to a new state.

\paragraph{Results:} Figure~\ref{fig:Main}B shows SFEC controlling the single SMD system toward a target position. 
The controller drives the mass on an exponential convergence to the target with minimal ringing, maintaining accurate tracking even under noise. 
The spike raster demonstrates sparse activity—spikes occur primarily during initial approach and when switching targets, with minimal activity once the target is reached. 
Fig.~\ref{fig:CoupledOscillatorAndDrones}A shows similar performance on the coupled oscillator system, where SFEC successfully coordinates the motion of multiple connected masses toward their respective targets (Fig.~\ref{fig:CoupledOscillatorAndDrones}A ii). 
In both cases, the Mean-Squared Error (MSE) and Free Energy remain low throughout the control episode, confirming that SFEC can stabilize and control oscillatory systems with multiple coupled degrees of freedom.

\vspace{2cm}
\subsection*{Extending to flexible target dynamics: drone swarm coordination}
\begin{figure}[hbtp!]
    \centering
    \includegraphics[width=\linewidth]{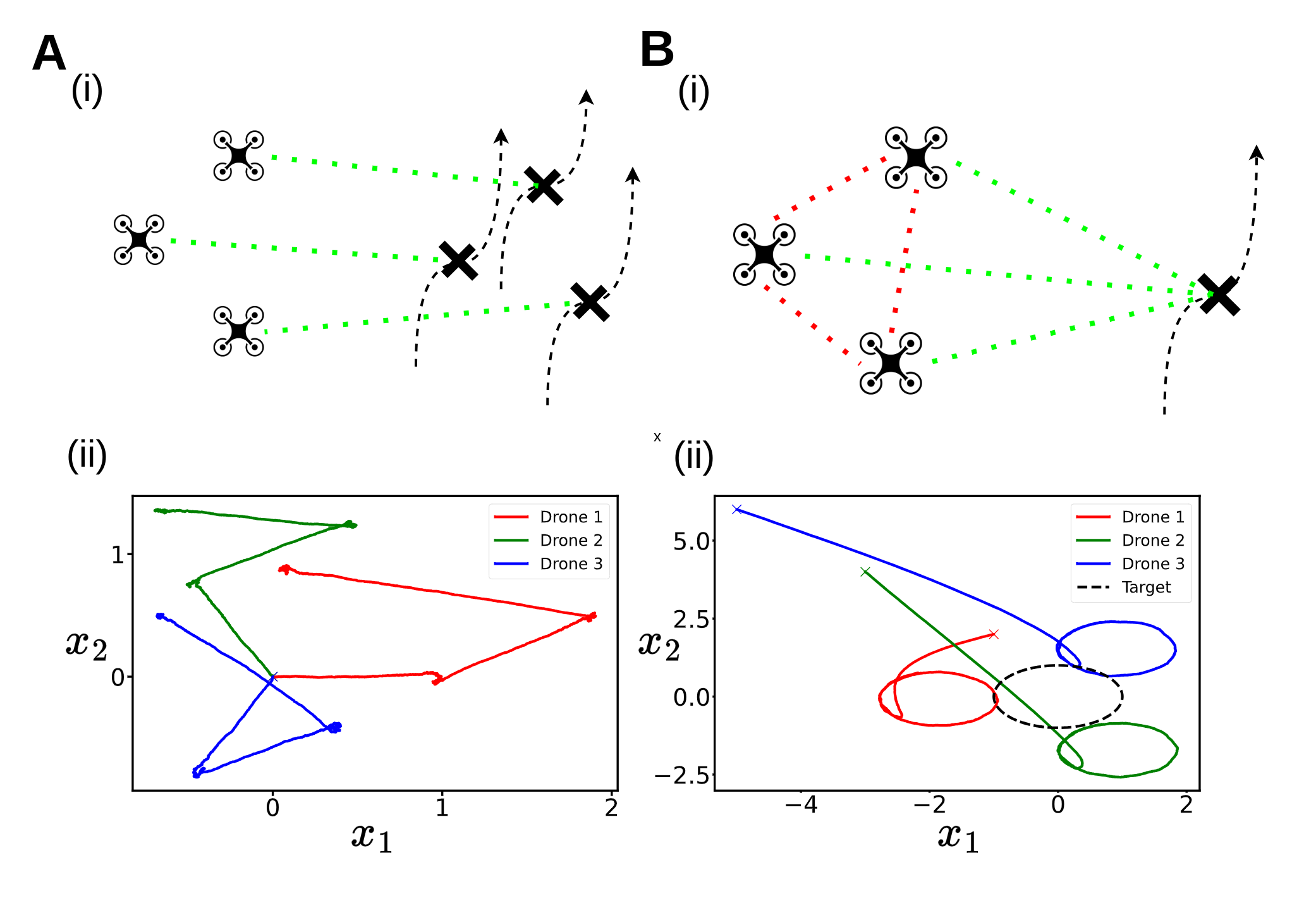}
    \caption{Flexible target dynamics enable different coordination behaviors.
    A) (i) Simple target dynamics: each mass is independently attracted to a target position (illustrated by springs from targets to drones, as illustrated by the green dotted lines); [explain green lines, arrows, X's].
    (ii) 2D trajectories of three drones following independent attractors — each drone moves directly to its target.
    B) (i) Formation target dynamics: masses are attracted to targets while repelling each other (green lines show attraction, red lines show repulsion).
    (ii) 2D trajectories showing formation maintenance — drones coordinate their motion to avoid collisions while following a central target which moves on a circle (dashed line).
    }
    \label{fig:DifferentDynamics}
\end{figure}

\paragraph{Problem:} While point-to-point control is useful, many real-world applications require coordinated multi-agent behaviors where agents should maintain specific spatial relationships while moving toward goals. 
For example, aerial drones performing surveillance or delivery tasks often need to maintain formations to optimize coverage or avoid collisions. 
Can SFEC be extended beyond simple target tracking to encode coordination behaviors?
\paragraph{Approach:} The key insight is that we minimize prediction errors not just about sensory observations, but also about dynamics. 
This allows us to specify desired system behaviors by defining a ``target dynamics'' model that the controller tries to make the real system follow. 
In the basic formulation used so far, this target dynamics is simply an attractor towards a fixed point. 
However, we have the freedom to specify any linear dynamical system as the target dynamics, enabling more sophisticated behaviors.
We demonstrate this flexibility on a Drone Swarm system consisting of 2D free masses, each with mass $m$ experiencing air friction coefficient $c$ (Fig.~\ref{fig:CoupledOscillatorAndDrones}B). 
We implement two distinct target dynamics:
\begin{enumerate}
    \item \textbf{Independent control} (Fig \ref{fig:CoupledOscillatorAndDrones}B, Fig.\ref{fig:DifferentDynamics}A): Each drone is attracted independently to its own target position, modeled as individual SMD systems connecting each mass to its target.
    \item \textbf{Formation control} (Fig.\ref{fig:DifferentDynamics}B): All drones are attracted toward a single target position while maintaining repulsive forces between each other, creating an emergent formation behavior. This is achieved by adding spring-like connections between all drone pairs in the target dynamics model.
\end{enumerate}

\paragraph{Results:} We first consider the case of independently controlling three drones. Fig. \ref{fig:CoupledOscillatorAndDrones}B shows the basic control of three drones and the corresponding spiking activity. As for the coupled SMDs the drone states follow the target states well, and spiking activity remains sparse except during target position changes. Figure~\ref{fig:DifferentDynamics}A illustrates the resulting trajectories, showing that drones can be controlled to follow a independent target trajectories. 
Figure~\ref{fig:DifferentDynamics}B demonstrates the formation control mode—the drones maintain spatial separation while moving their collective centroid toward the target location. 
The trajectories curve around each other, reflecting the repulsive interactions encoded in the target dynamics. 
Importantly, SFEC achieves both behaviors using the same underlying network architecture and derivation; only the specification of target dynamics changes. 
This demonstrates how Active Inference's dual error structure, which minimizes errors about both observations and dynamics, naturally enables flexible, task-specific control behaviors without architectural modifications.

\subsection*{Robustness to external perturbations}
\begin{figure*}
    \centering
    \includegraphics[width=0.90\linewidth]{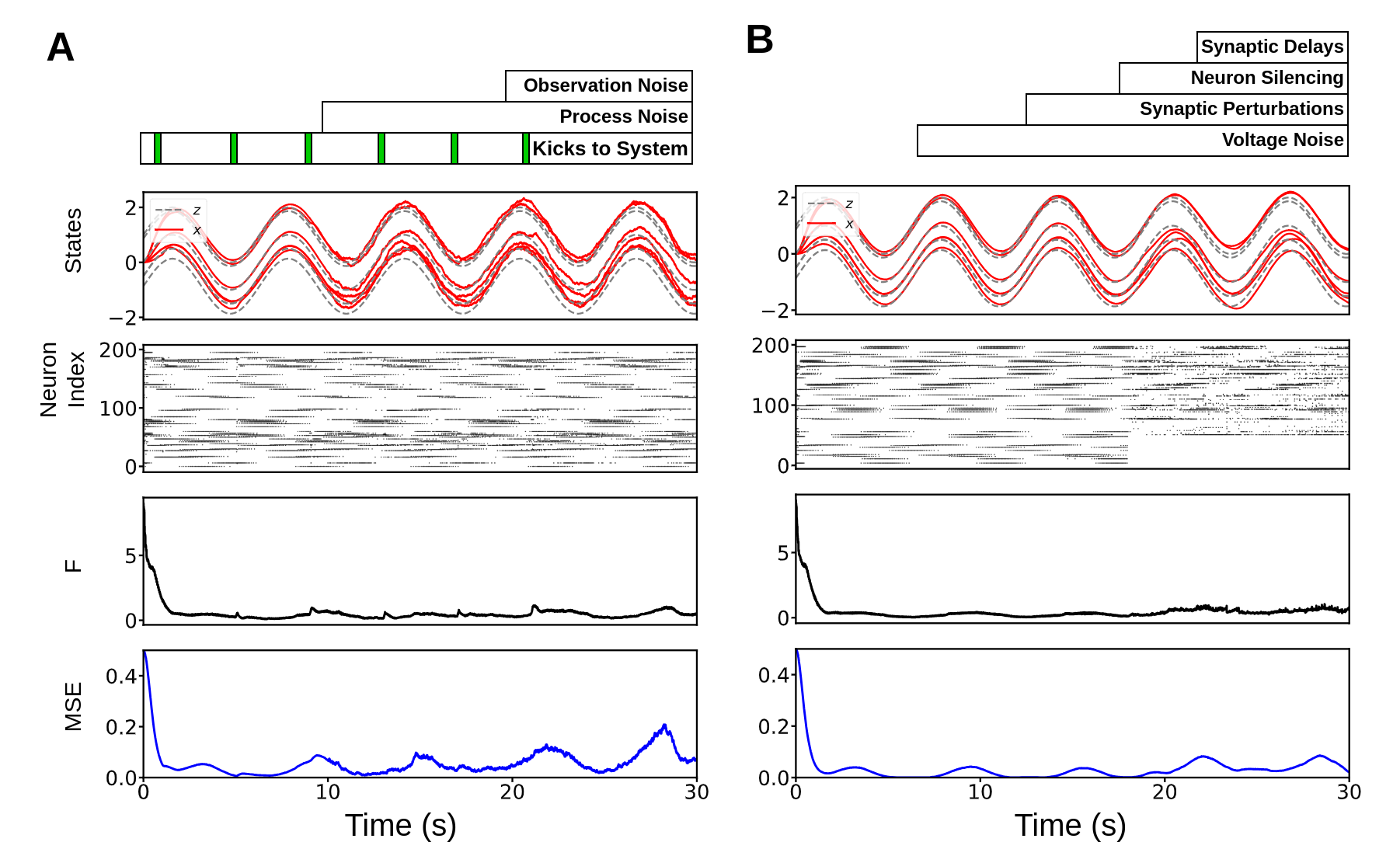}
    \caption{SFEC robustness to cumulative perturbations. (Top row) A timeline indicating when each perturbation is activated). (Rows 2-4) Target and state changes, spike raster, free energy, and the MSE.
    A) Control robustness under external perturbations, consisting of random kicks (throughout), increased process noise ($t \geq 10s$) and increased observation noise ($t \geq 20s$).
    B) Control robustness under internal perturbations applied to the controller network, consisting of voltage noise ($t \geq 6s$), synaptic perturbations ($t \geq 12s$), 25\% neuron silencing ($t \geq 18s$), synaptic delays ($t \geq 24s$).}
    \label{fig:Robustness}
\end{figure*}

\paragraph{Problem:} Real-world control systems must operate in uncertain environments with noisy sensors and unpredictable disturbances. 
How robust is SFEC to external perturbations that affect the plant dynamics and observations?
\paragraph{Approach:} We systematically stress-test SFEC on the drone swarm system by cumulatively introducing external perturbations (Fig.~\ref{fig:Robustness}A top).
We evolve the control episode with continuously varying targets instead of discreet jumps to better evaluate changes induced by the perturbations and show the state evolution and the spike raster, as well as the free energy and mean-squared-error metrics to evaluate performance:
\begin{itemize}
    \item \textbf{Robustness to process noise} ($t \geq 10s$): We increase the process noise by scaling the variance of the noise term in the dynamics update mechanism of the plant by a factor of $\sigma_{new} =100\sigma$.
    \item \textbf{Robustness to sensory noise} ($t \geq 20s$): We increase the sensory noise by scaling the variance of the noise term in the observation mechanism of the plant by a factor of $\sigma_{new} =100\sigma$.
    \item \textbf{Robustness to external kicks} ($t = (1+4n)s, n \in [0, 5]$): We add six kicks of magnitude $F_{kick} = 10$ in a random direction for a duration of $0.1s$ to the evolution of the system at periodic intervals.
\end{itemize}

\paragraph{Results:} Despite these perturbations, SFEC maintains stable control throughout the episode (Fig.~\ref{fig:Robustness}A). 
When random kicks occur, we observe brief spikes in both Free Energy and MSE as the system is perturbed away from targets, followed by rapid recovery as SFEC increases spike rate to correct the deviation (note the vertical lines appearing in the raster plot during the kicks). 
When process noise increases at $t=10s$, the baseline MSE rises modestly but control remains stable. 
Even after observation noise increases as well at $t=20s$, SFEC continues tracking the time-varying target, though with increased tracking error reflecting the fundamental information loss. 

\begin{figure}
    \centering
    \includegraphics[width=\linewidth]{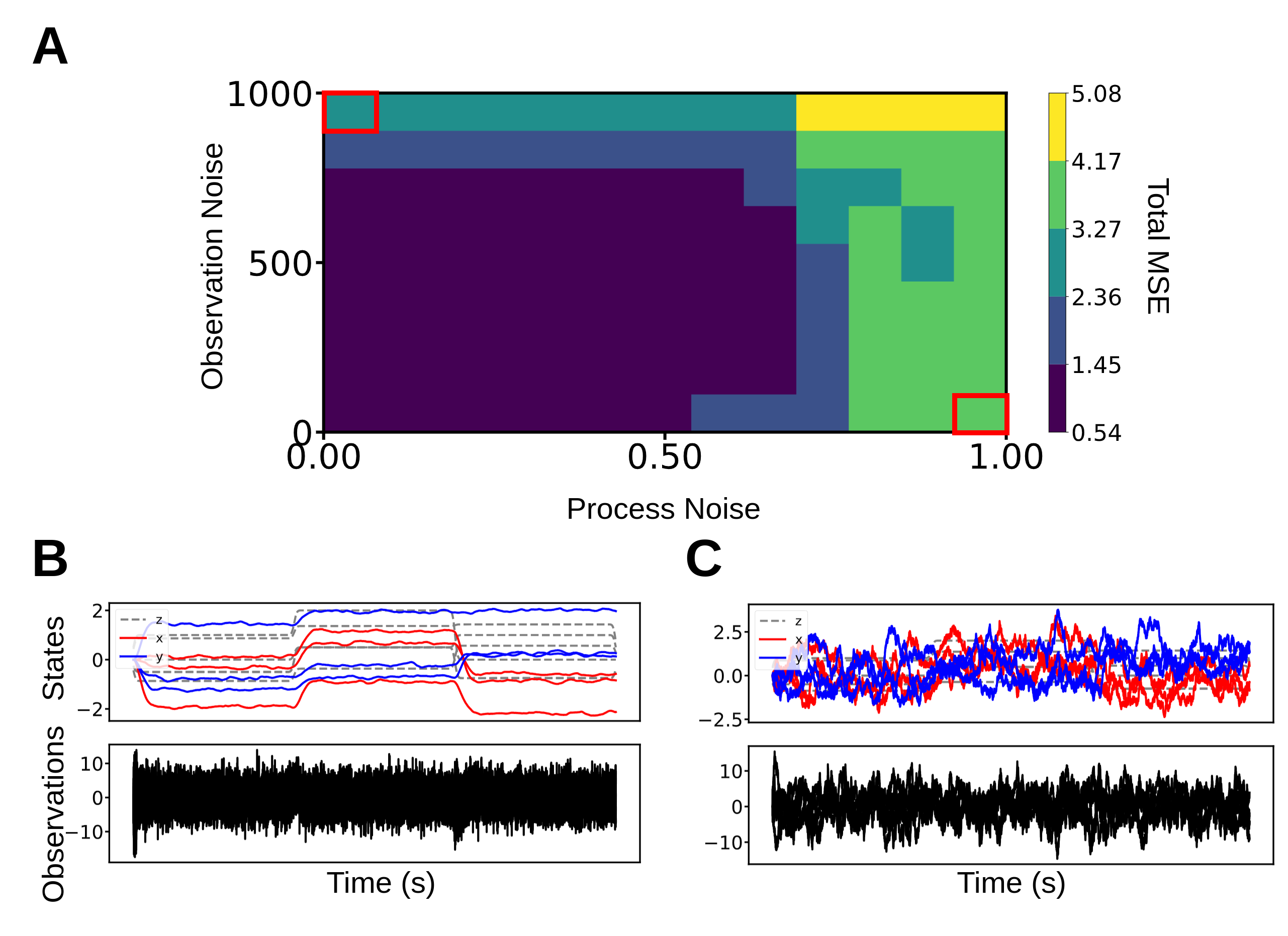}
    \caption{Extensive analysis of the controller's robustness to noise. 
    A) Heat Map plot showing Control Noise in the x-axis and Observation Noise in the y-axis; total MSE of the entire control run ($T=30s, dt=0.001s$, 2D masses) is shown as color, where brighter color represents more error.
    B) Example of a run with minimal Process Noise and very high Observation Noise, corresponding to the highlighted box on the top left.
    C) Example of a run with very high Process Noise and minimal Observation Noise, corresponding to the highlighted box on the bottom right.
    In both B and C, the top subplot shows the real positions of the masses, whilst the bottom subplot shows the observations fed to the controller.}
    \label{fig:Heatmap}
\end{figure}

To systematically characterize noise robustness, we perform a parameter sweep across process and observation noise levels (Fig.~\ref{fig:Heatmap}). 
The heatmap (Fig.~\ref{fig:Heatmap}A) shows MSE averaged over a full 30-second control episode across logarithmic ranges of both noise types. 
SFEC maintains low error across most of the space, degrading only in extreme high-noise corners. 
Figure\ref{fig:Heatmap}B-C shows example failure modes from opposing corners: with minimal process noise but extreme observation noise (Fig.~\ref{fig:Heatmap}B), the system state remains smooth but the controller cannot parse the real position from the noisy measurements; with extreme process noise but clean observations (Fig.~\ref{fig:Heatmap}C), the controller can ``see'' the real state of the system, but dynamics are noisy and uncontrollable.

\subsection*{Robustness to internal perturbations}
\paragraph{Problem:} Beyond external disturbances, biological neural systems and neuromorphic hardware experience internal failures: neurons die or become unreliable, synaptic weights drift, and signal transmission experiences delays.
For SFEC to serve as a model of biological robustness or as a practical neuromorphic controller, it must tolerate such internal perturbations. 
How does SFEC perform when its own computational substrate degrades?
\paragraph{Approach:} We cumulatively introduce internal network failures to the drone swarm controller (Fig.~\ref{fig:Robustness}B).
We also use continuously varying targets in this test and report the same metrics as for the external perturbations:
\begin{itemize}
    \item \textbf{Voltage noise} ($t \geq 6s$): Membrane potential noise variance increased 10-fold ($\sigma_{new} = 10\sigma$), simulating noisy neural integration.
    \item \textbf{Synaptic perturbations} ($t \geq 12s$): Each fast recurrent connection $\Omega_{fast,ij}$ is independently scaled by a random factor $a_{ij} \in [0.9, 1.1]$ at each time step, representing synaptic weight drift or manufacturing variability in neuromorphic chips.
    \item \textbf{Neuron silencing} ($t \geq 18s$): 25\% of neurons are permanently silenced, simulating cell death or hardware component failure.
    \item \textbf{Synaptic delays} ($t \geq 24s$): Spike transmission delay increased from one timestep to five timesteps (1ms to 5ms), representing slower or longer-range connections.
\end{itemize}

\paragraph{Results:} Remarkably, SFEC maintains control functionality under all internal perturbations (Fig.~\ref{fig:Robustness}B). 
Voltage noise at $t=6s$ causes a slight increase in spike rate but minimal change in MSE, as the population code absorbs the noise. 
Synaptic perturbations at $t=12s$ produce negligible effects—the network's redundancy allows it to compensate for ±10\% variations in connection weights. 
Most strikingly, even after silencing 25\% of neurons at $t=18s$, control performance degrades gracefully rather than catastrophically: MSE increases but the system continues tracking targets. 
The remaining 75\% of neurons simply spike more frequently to compensate for the lost capacity. 
Finally, introducing synaptic delays at $t=24s$ again increases spike rate and MSE moderately, but control remains stable. 
This graceful degradation under internal failures demonstrates that SFEC inherits the robustness properties of Spike Coding Networks \citep{calaim_geometry_2022}, where population-level representations naturally tolerate single-unit failures.

\subsection*{Comparison with alternative spiking implementations}
\paragraph{Problem:} While SFEC shows effective control and robustness, there are other ways to implement active inference in spiking neural networks. How does SFEC compare to other such methods? 
Specifically, is the direct free energy constraining approach more efficient than embedding gradient-based Active Inference into general-purpose spiking frameworks?
\paragraph{Approach:} To our knowledge there is no other spiking approach that directly minimizes or constraints the free energy. However, the classic approach of gradient descent on the free energy can be implemented in general spiking frameworks. We implemented this in two alternative approaches and compared them to SFEC on the drone swarm control task:
\begin{enumerate}
\item \textbf{NENGO}: Active Inference differential equations (obtained by gradient descent on the Free Energy landscape) implemented in the Neural Engineering Framework using the NENGO software package~\cite{eliasmith_neural_2004}.
\item \textbf{Gradient SCN}: The same Active Inference gradient equations implemented in a standard Spike Coding Network architecture~\cite{boerlin_predictive_2013}.
\end{enumerate}
For each method, we tested three population sizes (N=100, 200, 400 neurons) and measured three metrics over a 30-second control episode: Mean-Squared Error (MSE), total spike count, and computation time per timestep. 
We averaged results over 25 runs for each configuration. 
All controllers operated under identical conditions (same noise levels, targets, and timestep).

\begin{figure}[hbtp!]
\centering
\includegraphics[width=0.95\linewidth]{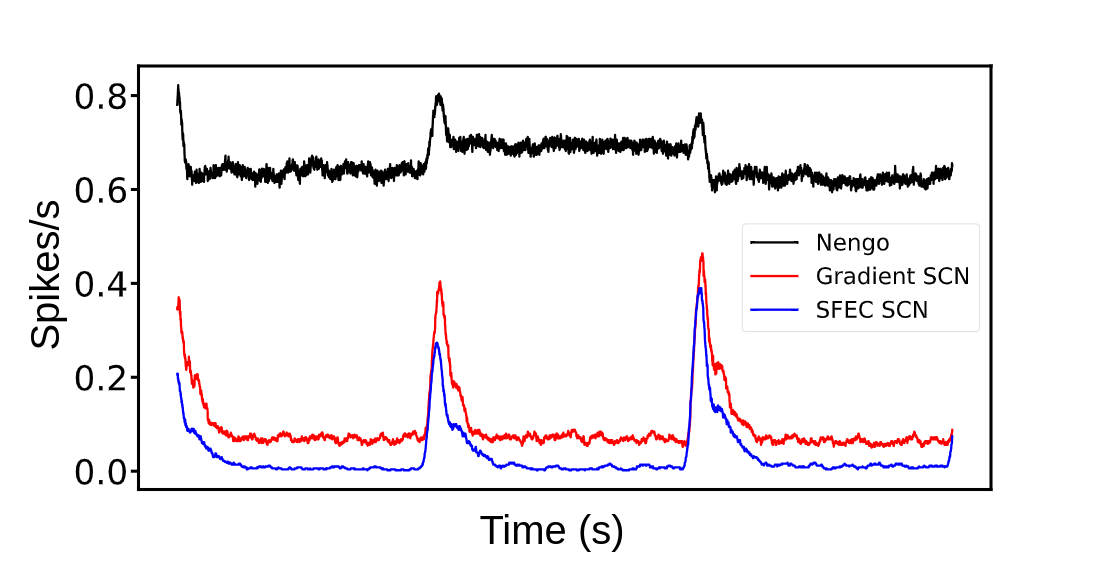}
\caption{\textbf{Total network firing rates over time.} 
Number of spikes per time step of each of the Spiking versions of the Free Energy Controller applied to the Drone Swarm system. 
Each firing rate is computed from it's corresponding network's spikes with a rolling window of 500 time steps. 
SFEC (blue) and Gradient SCN (red) are 1-2 orders of magnitude more efficient than the NENGO variant (black) in terms of spike counts, and SFEC displays the best performance, maintaining better efficiency than the Gradient SCN.}
\label{fig:EnergySpikesSCN}
\end{figure}

\paragraph{Results:} We can see in Table~\ref{tab:Table1} that SFEC is the most spike efficient of the implementations. 
This pattern of efficiency holds across all neuron counts: SFEC consistently uses 20-30x fewer spikes than NENGO and about 50 \% fewer than Gradient SCN, though NENGO does achieve lower MSE for higher neuron counts.
Figure~\ref{fig:EnergySpikesSCN} illustrates this efficiency difference dynamically, showing how all of the networks display increased activity during periods of high free energy when the target changes. SFEC, however, emits almost no spikes during periods when the target remains stationary.
The plot shows network-wide firing rates (spikes per second, computed with a 500-timestep rolling window) throughout the control episode. 
NENGO maintains high, constant firing rate regardless of task demands—evidence of the ``always-on'' nature of NEF population codes. 
The Gradient SCN reduces this substantially but still shows elevated baseline activity when compared to SFEC.
SFEC exhibits highly task-modulated spiking: firing rate spikes sharply when targets change (requiring rapid state updates), then drops nearly to zero during steady-state tracking. 
This directly reflects SFEC's design principle: spikes are fired only when they reduce Free Energy, naturally creating energy-efficient, adaptive computation.
Computation time per timestep remains comparable across methods (0.02-0.11ms), indicating that SFEC's efficiency gains come from algorithmic improvements rather than implementation artifacts.

\begin{table}[hbtp!]
    \centering
    \resizebox{\linewidth}{!}{
    \begin{tabular}{|c c c m{2.5cm}|} 
     \hline
     Controller (\#Neurons) & MSE & Spike Count & Comp. Time (ms/timestep)  \\ [0.5ex] 
     \hline\hline
     Nengo (100)  & $(1.90 \pm 0.22)e4$ & $11608 \pm 1209$ & 0.06  \\ 
     \hline
     Gradient SCN (100) &  $(1.68 \pm 0.01)e4$ & $2032 \pm 131$ & 0.02 \\
     \hline
     SFEC (100) &  $\mathbf{(1.60 \pm 0.02)e4}$ & $\mathbf{642 \pm 7}$ & 0.02 \\ [1ex] 
     \hline\hline
     Nengo (200)  & $\mathbf{(1.52 \pm 0.04)e4}$ & $19897 \pm 1904$ & 0.06  \\ 
     \hline
     Gradient SCN (200) &  $(1.66 \pm 0.01)e4$ & $2915 \pm 71$ & 0.03  \\
     \hline
     SFEC (200) & $(1.57 \pm 0.03)e4$ & $\mathbf{1053 \pm 30}$ & 0.04 \\ [1ex] 
     \hline\hline
     Nengo (400)  & $\mathbf{(1.45 \pm 0.02)e4}$ & $37138 \pm 2930$ & 0.06 \\ 
     \hline
     Gradient SCN (400) &  $(1.64 \pm 0.01)e4$ & $3850 \pm 144$ & 0.11  \\
     \hline
     SFEC (400) & $(1.53 \pm 0.01)e4$ & $\mathbf{1923 \pm 138}$ & 0.12\\ [1ex] 
     \hline
    \end{tabular}
    }
    \caption{\normalfont Performance and efficiency evaluation: Mean-Square-Error, Total Spike Count and Computation time of each of the implementations averaged over 20 runs, under the same conditions as all other tests ($T=30s, dt = 0.001s$), controlling the 2D masses system. Best values in bold and standard deviations reported.}
    \label{tab:Table1}
\end{table}

\section*{Conclusion}
We set out to answer a fundamental question: how can we build spiking control systems that exhibit the robustness to internal and external perturbations observed in biological brains? Our answer unifies two previously separate levels of analysis—the computational principles of Active Inference and the implementation substrate of spiking neural networks—by deriving spike-based dynamics directly from the variational free energy functional.
The standard approach to implementing Active Inference in spiking networks would be to first derive a solution using gradient descent on the variational free energy landscape, then embed this continuous-valued algorithm into a general-purpose spiking framework, such as the NEF or standard SCN. While this two-stage approach can work, it treats the spiking implementation as merely a substrate for executing an algorithm designed for continuous computation. Within Marr's levels of analysis~\citep{marr_vision_2010}, this nicely separates the algorithmic and implementation levels, which aids in interpretability and generality. However, this separation limits efficiency: the resulting networks must constantly maintain population representations even when the system state is stable, leading to unnecessary spike expenditure~\ref{fig:EnergySpikesSCN}.
Instead, we unified the algorithmic and implementation levels by deriving a fundamentally spike-based solution to Active Inference from first principles. By applying the error-constraining methodology of SCNs directly to the variational free energy expression—rather than to its gradient—we obtained the Spiking Free Energy Constrainer (SFEC), a network where each spike is guaranteed to reduce Free Energy. This principled derivation generalizes SCN voltage dynamics beyond simple signal-distance objectives to the weighted, precision-modulated prediction errors characteristic of Active Inference, resulting in a controller that is both biologically grounded and event-driven.
This unification yields performant practical advantages. Our experiments showed that SFEC achieves approximately $20$-fold greater spike efficiency compared to NENGO-based implementations while maintaining competitive control accuracy. Compared to gradient-based SCN implementations of Active Inference, SFEC still achieves 2-3× spike reductions for small networks, though efficiency converges between them as network size increases. This efficiency emerges naturally from SFEC's design: spikes occur only when needed to constrain errors, creating task-modulated activity that concentrates during state transitions and nearly vanishes during steady-state tracking. Furthermore, SFEC exhibits robust performance: it tolerates high levels of sensory and process noise, adapts to external perturbations, sustains functionality under synaptic weight perturbations, and maintains performance even when 25\% of neurons are silenced, demonstrating graceful degradation rather than catastrophic failure.

Beyond efficiency and robustness, our framework naturally accommodates flexible target dynamics. By encoding desired behaviors as dynamical systems that the controller makes the real system emulate, we demonstrated both independent target tracking and coordinated formation control in small drone swarms using identical network architecture. This flexibility stems directly from Active Inference's prediction error structure, which treats dynamics predictions on equal footing with sensory predictions.

\paragraph{Impact}
This work has implications across multiple domains. For robotics and neuromorphic engineering, SFEC provides a path toward more autonomous, efficient, and adaptive systems. The attractive spike efficiency positions SFEC as particularly suitable for energy-constrained neuromorphic hardware, where every spike incurs real power costs. As we transition toward edge computing and embodied AI systems, algorithms that naturally exploit event-driven computation will become increasingly valuable.
From a neuroscience perspective, this work offers a computational model that bridges theoretical principles with mechanistic implementation, providing a concrete hypothesis for how the brain might implement efficient motor control. By showing that the free energy principle can be directly instantiated in integrate-and-fire dynamics without intermediate gradient computations, we suggest a potential mechanism by which biological neural circuitry could perform complex control tasks with the robustness and efficiency observed in living organisms.
\paragraph{Limitations and future work}
Several extensions are necessary to fully realize SFEC's potential. First, empirical validation on physical robots~\cite{lanillos_active_2021} and low-power neuromorphic hardware~\cite{kudithipudi_neuromorphic_2025} remains essential to demonstrate real-world feasibility and quantify practical advantages over classical controllers. Prior~\cite{jimenez-fernandez_neuro-inspired_2012,slijkhuis2023closed} and ongoing~\cite{agliati_spiking_2025, casanueva-morato_space-time_2025} work has and is addressing these questions with related or similar approaches and algorithms.
Second, our current implementation is restricted to linear dynamical systems. Extending SFEC to non-linear systems will require either linearization approaches or modifications to the neuron model. Promising directions include neurons with richer dynamics~\cite{izhikevich_resonate-and-fire_2001}, excitory-inhibitory neuron interplay~\citep{podlaski_approximating_2024}, multi-compartment models with dendritic non-linearities~\cite{thalmeier_learning_2016, alemi_learning_2017, nardin_nonlinear_2021}, or hierarchical architectures where multiple linear SFEC layers compose to handle non-linearities.
Third, the current derivation assumes fixed, analytically-derived weights. Introducing plasticity mechanisms would enable SFEC to adapt to changing system dynamics or learn from experience. Biologically-plausible learning rules such as Hebbian plasticity, equilibrium propagation~\cite{scellier_equilibrium_2017}, Lagrangian-based methods~\cite{pourcel_lagrangian-based_2025}, or Hamiltonian echoes~\cite{oconnor_training_2019} could potentially be integrated while preserving SFEC's efficiency properties, as well as recent work in the SCN literature~\citep{geirhos_shortcut_2020, mancoo_understanding_2020}.
Finally, constructing spiking world models~\cite{taniguchi_world_2023}—potentially incorporating spiking memory architectures—could enable SFEC to handle partially-observable environments and long-horizon planning tasks, extending its applicability to more complex real-world scenarios.
In summary, by deriving spiking dynamics directly from the free energy principle rather than embedding gradient-based solutions into spiking substrates, we have demonstrated that the computational and implementation levels need not be separate. This unification yields networks that are simultaneously efficient, robust, flexible, and biologically plausible—characteristics that have long seemed in tension but here emerge naturally from a single mathematical framework. SFEC thus represents not just a new controller, but a methodological template for designing spike-based systems that fully exploit the event-driven nature of neural computation.

\section*{Materials and Methods}
\subsection*{Control Problem}
We want to control a linear time-invariant dynamical system (LTI) with a SNN to follow a given target or target state $\vec{z} \in \RR^{K}$. We assume the controller receives noisy observations $\vec{y}\in\RR^{K}$ of the current system state $\vec{x}\in\RR^{K}$ and actuates on the system by generating a control signal $\vec{u}\in\RR^{K/2}$.
$$
    \dot{\vec{x}} = \mathbf{A}\vec{x} + \mathbf{B} \vec{u} + d\vec{w}
$$
$$
    \vec{y} = \mathbf{C}\vec{x} + d\vec{\eta}
$$
where $\mathbf{A}\in\RR^{K\times K}$, $\mathbf{B}\in\RR^{K/2 \times K}$, $\mathbf{C}\in\RR^{K\times K}$ matrices define, respectively, the system dynamics, the control signal dynamics and the observability of the plant (the plant encompasses the dynamics of the system’s evolution and observation, including noise terms).
The control signal is obtained as a function of the controller's internal state.
The plant, upon simulating the current time step of the world dynamics, provides the controller with a sensory measurement and receives from it the motor control signal that will be integrated into the next time step.

We defined three dynamical systems (plant) with different complexities: The Spring-Mass-Damper (SMD) system, the coupled SMDs and the drone swarms, implemented as free masses.

For the SMD system, we use the $\mathbf{A}$ matrix:
$$ \mathbf{A} = \begin{bmatrix} 0 & 1 \\ -k & -c \end{bmatrix} $$
For reproducibility we used a  $k=3$ and $c=1$ for all simulations.

For the Coupled SMD system, we use an $\mathbf{A}$ matrix of the following structure (in the $n_{masses}=3$ case):
\[
A = \begin{bmatrix}
0 & 0 & 0 & 0 & 1 & 0 & 0 & 0\\
0 & 0 & 0 & 0 & 0 & 1 & 0 & 0\\
0 & 0 & 0 & 0 & 0 & 0 & 1 & 0\\
0 & 0 & 0 & 0 & 0 & 0 & 0 & 0\\
-{k_1 - k_2} & {k_2} & 0 & 0 & -c & 0 & 0 & 0\\
{k_2} & -{k_2 - k_3} & {k_3} & 0 & 0 & -c & 0 & 0\\
0 & {k_3} & -{k_3 - k_4} & 0 & 0 & 0 & -c & 0 \\
0 & 0 & 0 & 0 & 0 & 0 & 0 & 0
\end{bmatrix}
\]
with $c=0.1, \vec{k}=[k_1, k_2, k_3, k_4]$ in which $k_n = 1 + \xi$, where $\xi$ is a sample taken from an exponential probability distribution.
Note that the entries associated with the final ``wall" mass $m_4$ are set to zero such that this mass has no dynamics.

Finally, for drones, we have the $\mathbf{A}$ matrix:
\[
A = \begin{bmatrix}
\mathbf{0} & \mathbf{1} \\
\mathbf{0} & -c\mathbf{1} 
\end{bmatrix}
\]
Where $\mathbf{0}$ and $\mathbf{1}$ are the $2n_{drones}$-dimensional empty matrix and identity matrix, respectively and $c = 0.5$.

\subsection*{Spiking Free Energy Constrainer}

The main derivation of the SFEC is grounded on two main pillars of theory behind it, both of which are in some way ``descended" from the theory of Predictive Coding \cite{boerlin_predictive_2013, friston_free_2006}.
On one side, Active Inference offers an algorithmic framework for perceiving and acting upon the world that is based on the free energy principle, on the other hand, Spike Coding Networks are a mechanistic theory of how spiking neural networks should form connections in order to minimize specified constraints.

\subsubsection*{Active Inference}
\label{Methods:ActiveInference}
The Active Inference framework casts both perception and action as logical consequences of a system minimizing it’s internal free energy  \cite{friston_action_2010}. This is equivalent to estimating the probability distribution of the system’s states conditional on observations $p(x|y)$ through approximation by an internal tractable probability distribution, which is called the variational distribution $q(x)$ \cite{buckley_free_2017}. By defining the variational free energy $F$ as
$$
    F = \int q(\vec{x}) \ln{\frac{q(\vec{x})}{p(\vec{x} , \vec{y})}} d\vec{x}
$$
we can approximate the true posterior through the minimization of $F$:
$$
    q(\vec{x}) \approx p(\vec{x} | \vec{y}) \implies \min F 
$$
By assuming q(x) is a sharply peaked Gaussian distribution centered on $\vec{\mu}$, a vector of generalized coordinates representing the internal belief of the controller, and it’s dimensions are conditionally independent \cite{buckley_free_2017}, F can be approximated by
$$
    F = - \ln{p(\hat{\vec{x}}, \vec{y})}
$$
Using the Laplace assumption \cite{friston_variational_2007}, and thus considering Gaussian probability distributions, we can write

\begin{equation}
    F = \frac{1}{2}(\vec{\epsilon_y}^T \mathbf{P_y} \vec{\epsilon_y} + \vec{\epsilon_{\mu}}^T \mathbf{P_{\mu}} \vec{\epsilon_{\mu}}) + \log{\mathbf{\Sigma_{y}}} + \log{\mathbf{\Sigma_{\mu}}}
    \label{eq:VFEerror}
\end{equation}
with sensory prediction errors $\vec{\epsilon_y} = \vec{y} - \vec{g}(\hat{\vec{x}})$ and dynamics prediction errors $\vec{\epsilon_{\mu}} = \vec{f}(\hat{\vec{x}}, \vec{z}) - \vec{\mu}$, in which $\vec{g}(\hat{\vec{x}})$ is the generative model of the observations and $\vec{f}(\hat{\vec{x}}, \vec{z})$ is the generative model of the target state attractor dynamics, $\mathbf{P_y}$ and $\mathbf{P_z}$ are the precisions of each prediction error (inverse variances) and the $\mathbf{\Sigma}$s are their corresponding co-variance matrices (inverse precisions).
By setting the internal state to be a concatenation of the internal system estimate and the desired target $\vec{\mu} = [\hat{\vec{x}}, \vec{z}]\in\RR^{2K}$ and assuming $g$ and $f$ are linear functions on their arguments, we can re-write
\begin{equation} \vec{\epsilon_y} = \vec{y}^{+} - \mathbf{C}^{+}\vec{\mu}
\label{eq:yerror}
\end{equation}
\begin{equation}
\vec{\epsilon_{\mu}} = \mathbf{M}\vec{\mu} - \vec{\mu}
\label{eq:muerror}
\end{equation}

where we are assuming that our target variables are fed to the controller as observations $\vec{y}^{+} = [\vec{y}, \vec{z}]\in\RR^{2K}$, $\vec{C}^{+}\in\RR^{2K\times 2K}$ and the $\mathbf{M}\in\RR^{2K\times 2K}$ matrix maps the target variables to the internal space ($\mathbf{M}\vec{\mu}$ would be a vector that looks like $[\vec{z}, \vec{z}]$ and $\vec{\epsilon_{\mu}}$ would look like $[\vec{z} - \hat{\vec{x}}, \vec{0}]$).

Finally, we can compute the action to perform as a linear operation on the internal state by applying the target state attractor on the real system:
$$ \vec{u} = \mathbf{U} \vec{\mu} $$
$$ \vec{u} = \mathbf{B}^{-1} \mathbf{A}_{target} \vec{\mu} $$
where $\mathbf{A}_{target}\in\RR^{K\times 2K}$ is the preferred dynamics matrix that the controller is making the real system follow. For instance, we can chose this desired dynamics as a set of springs connecting the masses to their target positions and giving them some damping to prevent ringing, as described in the result of the drones with different formation dynamics (Fig. \ref{fig:DifferentDynamics}). The structure and parameters of $\mathbf{A}_{target}$ is specified in the Supplementary Materials.

Now we have the ingredients for a non-spiking algorithm implementing active inference to control LTI dynamical systems. In the following, we describe SCNs and how to specify the internal dynamics of the SNN to constraint the free energy $F$.

\subsubsection*{Spike Coding Networks}
\label{Methods:SCN}
The neuroscience theory of SCNs offers a robust and efficient method of encoding linear dynamical systems into populations of spiking neurons \cite{boerlin_predictive_2013}. The SCN formalism allows this by maintaining an internal representation of the signal being tracked and constraining its error to remain within certain bounds - every time this error crosses a certain threshold, a neuron will spike to bring the representation back within bounds. This provides tolerance to various types of disturbances, ranging from synaptic perturbations and delays to total neuron silencing \cite{calaim_geometry_2022}. Any linear controller can be instantiated within an SCN and obtain these robustness and efficiency benefits---see \cite{slijkhuis2023closed} for the spiking version of the linear quadratic gaussian controller.

The SCN framework has also been shown to be capable of maintaining proper signal representations with fewer neurons than NEF networks, to be more efficient in terms of sparsity (the neurons present are less active than in NEF networks), and to be orders of magnitude more reliable despite exhibiting similar single-unit properties~\cite{boerlin_predictive_2013, deneve_efficient_2016}. Furthermore, SCNs are remarkably robust to network perturbations such as voltage noise, synaptic weight perturbations, transmission delays, and even neuron loss, due to their balanced excitatory-inhibitory dynamics and the geometric properties of their representation space~\cite{calaim_geometry_2022}.

Within this framework, we assume the filtered spike trains of $N$ neurons $\vec{r}\in\RR^{N}$ encode the internal variable of the network $\vec{\mu}\in\RR^{K}$ through a linear decoding matrix $\mathbf{D}\in\RR^{N\times K}$:
$$
    \vec{\mu} = \mathbf{D}\vec{r}
$$
$$
    \dot{\vec{r}} = -\lambda \vec{r} + \vec{s}
$$
where $\lambda$ is the leak constant of the readout. The structure of $\boldsymbol{D}$ is specified in the Supplementary Materials for both the original SCN and the SFEC.

We can then define the loss function between this encoded variable and a target variable, and perform the optimization procedure described in \cite{boerlin_predictive_2013} that gives us the voltage equations. The classical SCN loss function is the quadratic error between the signal estimate and the target signal:
$$
    L(\vec{x}, \vec{\mu}) = ||\vec{x}-\vec{\mu}||^{2}_{2}
$$
which gives us voltage equation 
$$
\vec{v} = \vec{D}^T(\vec{x}-\vec{\mu})
$$
By modifying this quadratic loss function with an extra term:
$$ L(\vec{x}, \vec{\mu}) = ||\vec{x}-\vec{\mu}||^{2}_{2} + ||\vec{r}||^{2}_{2} $$ 
the voltage dynamics remain unchanged, but our model develops an adaptive threshold $T_i$ for each neuron $i$, which makes the network activity more sparse and distributed, and makes the network more tolerant to synaptic delays \cite{boerlin_predictive_2013, calaim_geometry_2022}:
$$ T_i = \frac{1}{2}\vec{D}_i^T\vec{D}_i + r_i + \frac{1}{2} $$

\subsection*{SFEC Derivation}
Here, we combine the active inference description for controlling LTI systems and the SCN framework to obtain the SFEC algorithm. To this end, instead of using the classical loss function of SCN we derive the new neuron spiking rule through the Loss function inequality using the free energy functional.
Given the variational free energy (Eq.~\ref{eq:VFEerror}) we add a $\mathbf{r^T r}$ term to the spike-based minimization process to guarantee sparsity and tolerance to delays,

$$ 
F = \mathbf{\epsilon_y}^T \mathbf{P_y}\mathbf{\epsilon_y} + \mathbf{\epsilon_{\mu}}^T \mathbf{P_{\mu}}\mathbf{\epsilon_{\mu}} + \mathbf{r^T r} + K
$$
where $K = \log{\mathbf{\Sigma_{y}}} + \log{\mathbf{\Sigma_{\mu}}}$ and the prediction errors are (from Eq. \ref{eq:yerror} and \ref{eq:muerror}),

$$ \vec{\epsilon}_y = \vec{y}^{+} - \mathbf{C}^{+}\vec{\mu} $$
$$ \vec{\epsilon}_{\mu} = \mathbf{M}\vec{\mu} - \vec{\mu} $$

We transform the error equations into matrix form by defining a vector $\vec{a} = [\vec{y}^{+}, \mathbf{M}\vec{\mu}]\in\RR^{4K}$ (where $\vec{y}^{+} = [\vec{y}, \vec{z}]\in\RR^{2K}$ and the $\mathbf{M}\vec{\mu}\in\RR^{2K}$ term is effectively the state if it were in the target position $\vec{x} = \vec{z}$) and a matrix $\mathbf{H} = \begin{bmatrix} \mathbf{C}^{+} \\ \mathbf{1} \end{bmatrix}\in\RR^{2K\times 4K}$ to get 
$$ \vec{\epsilon} = \vec{a} - \mathbf{H}\vec{\mu} \in\RR^{4K}$$
and re-write
$$ F = \mathbf{\epsilon}^T \mathbf{P}\mathbf{\epsilon} + \mathbf{r^T r} + K$$
Deriving the voltage equation and spiking rule can then be achieved by resolving the following inequality, for the $i$th neuron:
$$ F(i=spike) < F(i=silent) $$
$$ \vec{D}_i^T \mathbf{H}^T \mathbf{P}\vec{\epsilon} \geq \frac{1}{2}\vec{D}_i^T \mathbf{H}^T \mathbf{P}\mathbf{H} \vec{D}_i + r_i + \frac{1}{2} $$

We can then identify the left side of the inequality as the expression for the neuron voltages and the right hand side as the neuron thresholds 
$$ \vec{v} = \mathbf{D}^T\mathbf{H}^T\mathbf{P}(\vec{a} - \mathbf{H}\vec{\mu}) $$
$$ T_i = \frac{1}{2}\vec{D}_i^T \mathbf{H}^T \mathbf{P} \mathbf{H} \vec{D}_i + r_i + \frac{1}{2} $$
The dynamical voltage equation then takes the shape 
$$ \dot{\vec{v}} = \mathbf{D}^T\mathbf{H}^T\mathbf{P}(\dot{\vec{a}}) - \mathbf{D}^T\mathbf{H}^T\mathbf{P}\mathbf{H}(-\lambda\vec{\mu} + \mathbf{D}\vec{s}) $$
$$ \dot{\vec{v}} = \mathbf{D}^T\mathbf{H}^T\mathbf{P}(\dot{\vec{a}}) + \lambda\mathbf{D}^T\mathbf{H}^T\mathbf{P}\mathbf{H}\mathbf{D}\vec{r} - \mathbf{D}^T\mathbf{H}^T\mathbf{P}\mathbf{H}\mathbf{D}\vec{s} $$
Separating the $\vec{a}$ vector back into it's components, we get
$$ 
\dot{\vec{v}} = \mathbf{D}^T\mathbf{H}^T\mathbf{P}\mathbf{1_y}(\frac{d}{dt}\vec{y}^{+}) + \mathbf{D}^T\mathbf{H}^T\mathbf{P}\mathbf{1_{\mu}}(\frac{d}{dt}(\mathbf{M}\vec{\mu}))
$$
$$
+ \lambda\mathbf{D}^T\mathbf{H}^T\mathbf{P}\mathbf{H}\mathbf{D}\vec{r} - \mathbf{D}^T\mathbf{H}^T\mathbf{P}\mathbf{H}\mathbf{D}\vec{s}
$$
with $ \mathbf{1_{y}} = \begin{bmatrix} \mathbf{1}\\\mathbf{0} \end{bmatrix}$ and $\mathbf{1_{\mu}} = \begin{bmatrix} \mathbf{0}\\\mathbf{1} \end{bmatrix}$. \\
Note that the derivatives in the expression are ill-posed for an instantaneous formulation. During computation, we will calculate $\dot{\vec{y}}^{+}_t = (\vec{y}^{+}_{t} - \vec{y}^{+}_{t-1})/dt $ as a Zero-Order-Hold approximation. For the derivative of the $\mathbf{M}\vec{\mu}$ term, we will define this as the target dynamics of the system, and set them according to the behaviour we want to impart on the controller $\frac{d}{dt}(\mathbf{M}\vec{\mu}) = \mathbf{A}_{target}\vec{\mu}$. \\
We then get

$$
\dot{\vec{v}} = \mathbf{D}^T\mathbf{H}^T\mathbf{P}\mathbf{1_y}\dot{\vec{y}}^{+} + \mathbf{D}^T\mathbf{H}^T\mathbf{P}\mathbf{1_{\mu}}\mathbf{A}_{target}\mathbf{D}\vec{r} $$
$$
+ \lambda\mathbf{D}^T\mathbf{H}^T\mathbf{P}\mathbf{H}\mathbf{D}\vec{r} - \mathbf{D}^T\mathbf{H}^T\mathbf{P}\mathbf{H}\mathbf{D}\vec{s}
$$
or, collecting terms for clarity
$$
\dot{\vec{v}} = \mathbf{W}_y\dot{\vec{y}}^{+} + \mathbf{\Omega}_{slow}\vec{r} + \mathbf{\Omega}_{fast}\vec{s}
$$
with
$$ \mathbf{W}_y = \mathbf{D}^T\mathbf{H}^T\mathbf{P}\mathbf{1_y}$$
$$ \mathbf{\Omega}_{slow} = \mathbf{D}^T\mathbf{H}^T\mathbf{P}(\mathbf{1_{\mu}}\mathbf{A}_{target} + \lambda\mathbf{H})\mathbf{D} $$
$$ \mathbf{\Omega}_{fast} = - \mathbf{D}^T\mathbf{H}^T\mathbf{P}\mathbf{H}\mathbf{D} $$

This network will naturally constrain the Free Energy of the system as defined in the Active Inference Controller derivation, meaning that the control output can simply be computed as the same linear operation on the internal state $\vec{\mu}$
$$ \vec{u} = \mathbf{U} \vec{\mu} $$

\subsection*{Implementation of other methods compared}

\subsubsection*{Free Energy Gradient Descent}
\label{Methods:FreeEnergyGradientDescent}

One way to minimize the Free Energy in an Active Inference Controller is to perform Gradient Descent on the Free Energy landscape \cite{pezzato_novel_2020}, leading to the following derivation:
$$ \dot{\vec{\mu}} = - \eta \frac{\partial F}{\partial \vec{\mu}}$$
the gradient of the free energy can be separated the gradients of the constituent errors
$$ \dot{\vec{\mu}} = - \eta (\frac{\partial \vec{\epsilon}_y}{\partial \vec{\mu}}^T\vec{\epsilon}_y + \frac{\partial \vec{\epsilon}_{\mu}}{\partial \vec{\mu}}^T\vec{\epsilon}_{\mu}) $$
replacing with $\vec{\epsilon}_y = \vec{y}^{+} - \mathbf{C}^{+}\vec{\mu}$ and $ \vec{\epsilon}_{\mu} = \mathbf{M}\vec{\mu} - \vec{\mu} $
$$ \dot{\vec{\mu}} = \eta (\mathbf{C^{+}}^T\mathbf{P}_y(\vec{y}^{+} - \mathbf{C}^{+}\vec{\mu}) + (\mathbf{1}-\mathbf{M})^T\mathbf{P}(\mathbf{M} - \mathbf{1})\vec{\mu}) $$
$$ \dot{\vec{\mu}} = \eta \mathbf{C^{+}}^T\mathbf{P}_y\vec{y}^{+} + \eta (-\mathbf{C^{+}}^T\mathbf{P}_y\mathbf{C}^{+}\vec{\mu} + (\mathbf{1}-\mathbf{M})^T\mathbf{P}(\mathbf{M} - \mathbf{1}))\vec{\mu} $$
Simplifying
$$ \dot{\vec{\mu}} = \mathbf{O}_y\vec{y}^{+} + \mathbf{O}_{\mu}\vec{\mu} $$
$$ \mathbf{O}_y = \eta \mathbf{C^{+}}^T\mathbf{P}_y $$
$$ \mathbf{O}_{\mu} = \eta (-\mathbf{C^{+}}^T\mathbf{P}_y\mathbf{C}^{+} + (\mathbf{1}-\mathbf{M})^T\mathbf{P}(\mathbf{M} - \mathbf{1})) $$
These are the dynamics encoded into the NENGO and Gradient SCN implementations against which we compare our algorithm.

\subsubsection*{Neural Engineering Framework (NEF)}
\label{Methods:nengo}
The NEF provides a method of performing such an implementation based on 3 core principles: Representation, Transformation and Dynamics \cite{eliasmith_neural_2004}, allowing the development of networks that implement different signals, linear or non-linear transformations and dynamical systems within the connectivity and parameters of a network. The NEF is implemented in the popular software package NENGO \cite{bekolay_nengo_2014}, which can support rate-based or Spiking Neurons with different neuron models and which calculates the initialization of network parameters by adjusting the tuning curves of individual neurons. 
We used NENGO to implement directly the set of dynamical equations derived for the controller from the Free Energy gradient descent.

\subsubsection*{SCN}
The SCN controller implements the dynamical equations derived for the Free Energy controller through Gradient Descent, using parameters detailed in the Supplementary Materials.
They are embedded into the network by assuming the derivative of the signal being encoded is given by the recurrent connections \cite{boerlin_predictive_2013}.

\subsection*{Simulation Parameters}
Numerical simulations of the system plant are computed with a Runge-Kutta-4 integration technique to guarantee fidelity and the Controller operates as a Spiking Neural Network.
In all simulations of the control problem, we set the time step to be $dt = 0.001s$ and the total duration of the experiment $T =30s$. We also inject process and observation noise from Gaussian distributions with variance $\sigma = 1\times 10^{-4}$ unless specified differently. 

\subsection*{Supporting Information Appendix (SI)}
We provide an extra comparison of the 3 different controllers, as well as some parameter definitions in the Supplementary Material.
All code used to simulate systems and generate plots available at: under review, ask the authors.

\subsection*{Acknowledgment}
ARU is funded by the JAE program (JAEPRE23-24) Spanish National Research Council. This work was partially supported by the Spikeference project, Human Brain Project SGA3 (Grant Agreement Nr. 945539).

\bibliographystyle{abbrvnat}
\bibliography{Paper.bib}

\section{Supplementary Materials}
\subsection{SFEC and SCN Decoders}
The decoder matrix $\mathbf{D}$ for both the Spiking Free Energy Constrainer (SFEC) and the Gradient SCN is constructed to ensure that the network can effectively span the state space while maintaining the bounding box mechanism. The initialization strategy involves creating a structured basis set for the first $2K$  neurons (where $K$ is the dimensionality of the internal state estimate $\mathbf{\mu}$) and filling the remaining neurons with random directions. The columns are then randomly permuted to distribute the specific functional roles across the neuron population.
For the Gradient SCN, we scale all decoders by diving their magnitude by the total number of neurons $N$ and multiplying by a scaling factor $\mathbf{D} = \frac{m}{N} \mathbf{D}_{unscaled}$.
For the SCN and SFEC in all cases but one, we chose $m=10$, however for SFEC control on the SMD system, we found better results with $m_{SMD} = 1$.

\subsection{Target Dynamics Parameters}
Each system being controlled requires that a Target Dynamics be defined for it.
For the Spring-Mass-Damper and for the coupled Spring-Mass-Damper systems, we use as the target dynamics another set of SMD systems attached between each mass and each target position, with spring constant $k=10$ and damping strength $c=5$.
For the Drone Swarm system, we use two different controllers with different target dynamics. In one, each drone is connected to each target by an SMD system (Independent control), whereas in the other, each drone is connected to each other drone repulsively and they are all connected attractively to a single target position (Coordinated control). In both cases the spring constant $k=5$ and the damping strength $c=5$ are the same.
In the case of Coordinated control, we also add a fixed term to the action $u$ which sets the equilibrium position of the formation.

\section{Supplementary Results}
\subsection{Comparison of controllers}
\begin{figure*}[hbtp!]
    \centering
    \includegraphics[width=0.9\linewidth]{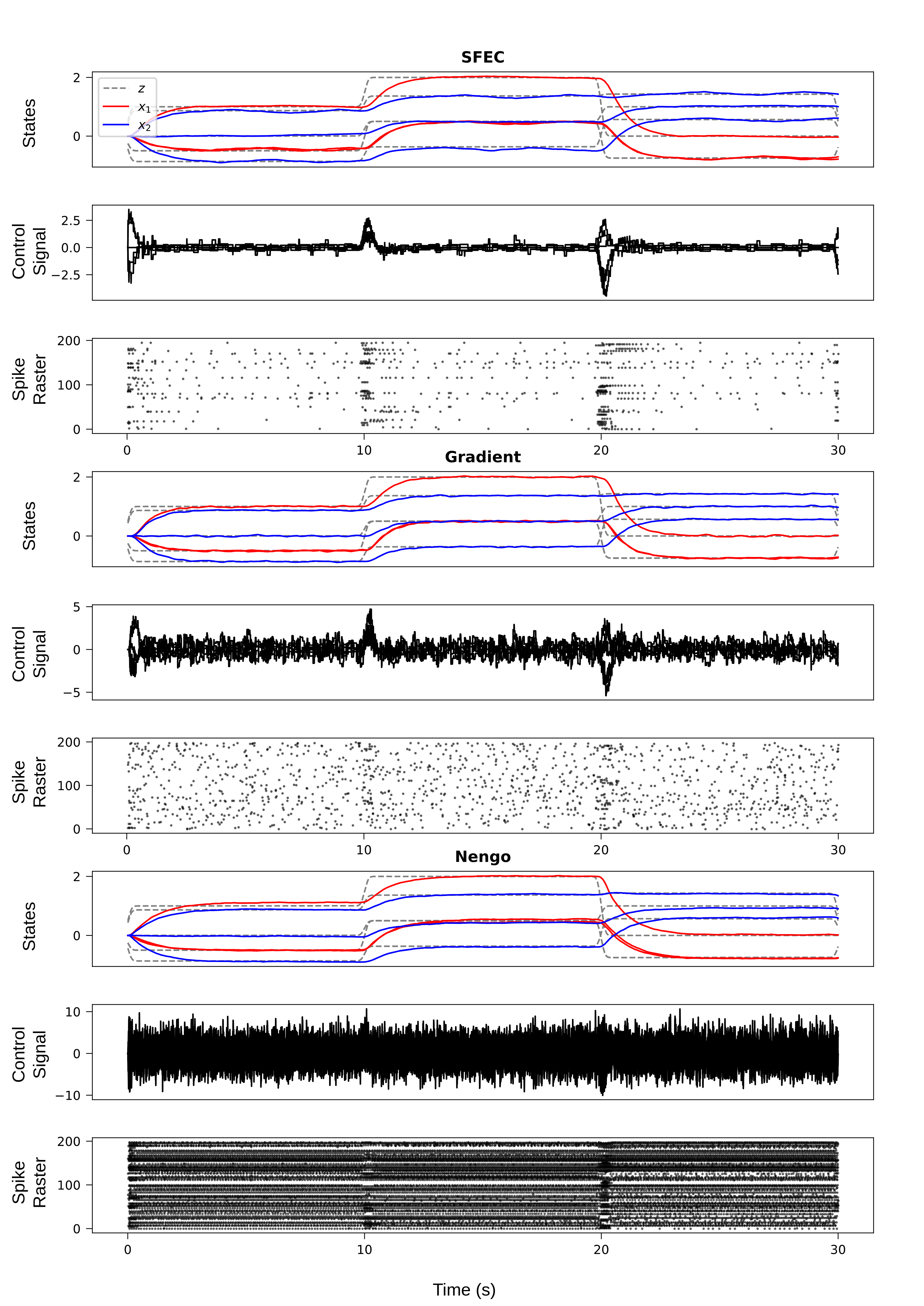}
    \caption{Comparison of the 3 controllers operating on the Drone Swarm system with 200 neurons and fixed conditions equal to Fig. 2B. Reporting the states, the generated control signal and the spiking activity.}
    \label{fig:placeholder}
\end{figure*}

\end{document}